\documentclass[journal=jacsat,manuscript=article,layout=twocolumn]{achemso}

\makeatletter
\let\l@addto@macro\relax
\makeatother
\usepackage[fontsize=11pt]{scrextend}
\usepackage[colorlinks,citecolor=red,urlcolor=blue,bookmarks=false,hypertexnames=true]{hyperref} 
\usepackage{color,soul}
\usepackage{booktabs}
\usepackage{multirow}
\usepackage{graphicx}
\usepackage[version=3]{mhchem} 
\let\oldmaketitle\maketitle
\let\maketitle\relax
\author{Lorenzo Varrassi}
\affiliation{Dipartimento di Fisica e Astronomia, Università di Bologna, 40127 Bologna, Italy}
\email{lorenzo.varrassi3@unibo.it}
\author{Peitao Liu}
\affiliation{Shenyang National Laboratory for Materials Science, Institute of Metal Research, Chinese Academy of Sciences,
110016 Shenyang, Liaoning, China}
\author{Cesare Franchini}
\affiliation{Dipartimento di Fisica e Astronomia, Università di Bologna, 40127 Bologna, Italy}
\alsoaffiliation{University of Vienna, Faculty of Physics and Center for Computational Materials Science, Kolingasse 14-16, A-1090, Vienna, Austria}

\title[]{Quasiparticle and excitonic properties of monolayer SrTiO$_3$}
\begin{document}

\twocolumn[
\begin{@twocolumnfalse}
\oldmaketitle
\begin{abstract}
Strontium titanate SrTiO$_3$ is one of the most studied and paradigmatic transition metal oxides.
Recently, a breakthrough has been achieved with the fabrication of freestanding SrTiO$_3$ ultrathin films down to the monolayer limit.
However, the many-body effects on the quasiparticle and optical properties of monolayer SrTiO$_3$ remain unexplored.
Using state-of-the-art many-body perturbation theory in the GW approximation combined with the Bethe-Salpeter equation,
we study the quasiparticle band structure, optical and excitonic properties of monolayer SrTiO$_3$.
We show that quasiparticle corrections significantly alter the band structure topology; however, the
widely used diagonal $G_0W_0$ approach yields unphysical band dispersions.
The correct band dispersions are restored only by taking into account the off-diagonal elements of the self-energy.
The optical properties are studied both in the optical limit and for finite momenta by computing the electron energy loss spectra.
We find that the imaginary part of dielectric function at the long wavelength limit is dominated by three strongly bound excitonic peaks
and the direct optical gap is associated to a bright exciton state with a large binding energy of 0.93 eV.
We discuss the character of the excitonic peaks via the contributing interband transitions,
and reveal that the lowest bound excitonic state becomes optical inactive for finite momenta
along $\Gamma$-M, while the other two excitonic peaks disperse to higher energies and eventually merge for momenta close to M.
\end{abstract}
\end{@twocolumnfalse}
]
\begin{figure}[t]
    \centering
    \includegraphics[width=0.975\linewidth]{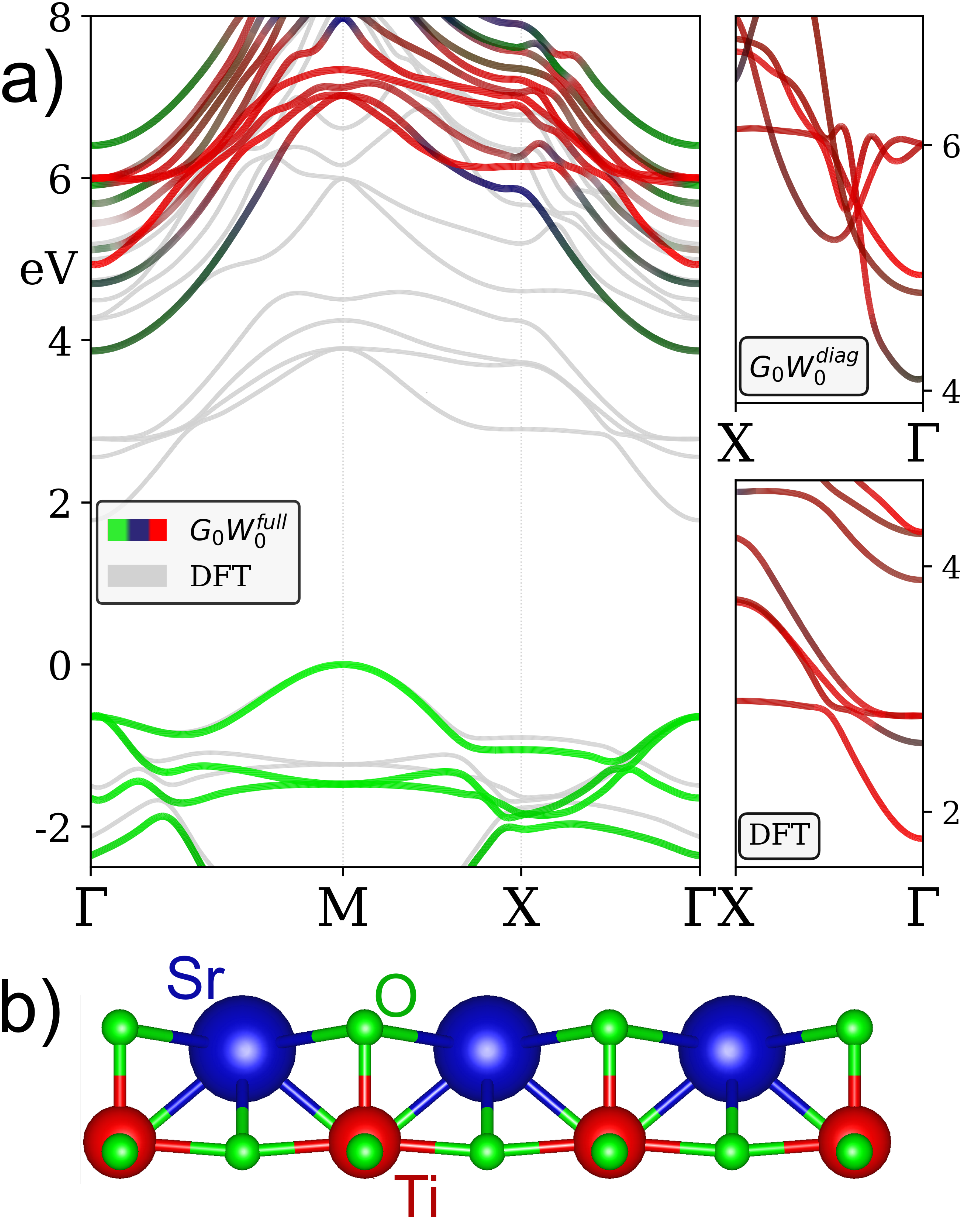}
	\caption{\label{fig:QP_KS_BS} a) DFT (grey lines) and orbital-projected $G_0W_0^{full}$ band structures. Colors represent the O, Ti and Sr characters respectively, The two right panels show the orbital-projected unoccupied bands along X-$\Gamma$ calculated by $G_0W_0^{diag}$ and DFT. b) Distorted monolayer SrTiO$_3$ structure.}
\end{figure}
Transition Metal Oxide (TMO) perovskites have attracted wide interest in the last years due to the many intriguing physical properties and possible technological applications in various fields as oxide electronics, spintronic or catalysis~\cite{Dogan,Guan2014, LIU2008701, SHOJI2016309}.
Among them SrTiO$_3$ has acquired a prototypical role: it's one of the most studied perovskites and has been employed as a proving ground to propose or compare different computational schemes~\cite{PhysRevB.85.245124,Ekuma_STO}, including many-body \textit{ab initio} methods~\cite{Giancarlo_Cappellini_2000,PhysRevMaterials.5.074601,PhysRevB.87.235102, PhysRevB.92.035119,PhysRevMaterials.2.024601} and machine-learning based algorithms~\cite{PhysRevMaterials.6.094408,Ranalli_future_work}.
The role of its electronic structure in determining the conducting, magnetic and optical properties has been widely investigated, and intriguing phenomena such as superconductivity or two dimensional electron gas~\cite{Santander-Syro2011-be,pnas.1318304111,Meevasana2011} have been identified and analyzed.
In particular recent theoretical works~\cite{PhysRevMaterials.5.074601,PhysRevB.87.235102,PhysRevMaterials.3.065004,PhysRevB.92.035119} analyzed the role of electronic correlations and localized $d$ states in the optical response, and highlighted how including an explicit description of excitonic interactions is necessary to achieve a satisfactory agreement with the experimental data.
\newline Recently, an important breakthrough has been achieved by Ji \textit{et al.}~\cite{Ji2019} through the synthesis of freestanding two-dimensional (2D) perovskites SrTiO$_3$ and BiFeO$_3$ films with a thickness reaching the monolayer limit.  Their work proves that TMO perovskites films can be realized with thicknesses below the critical limit previously proposed as necessary for crystalline order stability~\cite{ciadv.aao5173}.  A year later freestanding PbTiO$_3$ films were fabricated with thicknesses down to four unit cells by Han \textit{et al}~\cite{admi.201901604}.
\newline In the last decades low-dimensional materials have attracted increasing attention due to their remarkable physical properties resulting from quantum confinement and reduced dimensionality effects, which differentiate them from bulk phases.
In particular, 2D monolayer structures have been intensely studied due to their fascinating optical and excitonic physics: the enhanced electron-hole interactions result in strongly bound excitons which dominate the optical response and charge transfer properties.~\cite{BookHaug,
PhysRevLett.111.216805,PhysRevB.98.125206,PhysRevB.93.235435,PhysRevLett.103.186802,Fugallo_2021,PhysRevB.99.075421,PhysRevLett.116.066803,RevModPhys.90.021001,PhysRevB.88.245309,PhysRevLett.113.076802,PhysRevB.99.161201}	
\begin{table}[b!]
\begin{tabular}{lll}
\hline \hline
                                         & Direct (eV)       & Indirect (eV)     \\ \midrule
BSE@G$_0$W$_0$$^{full}$($E_{xc}$) & 3.588 (0.93) & 2.891 (0.98) \\
G$_0$W$_0$$^{full}$                      & 4.514        & 3.870        \\
G$_0$W$_0$$^{diag}$                      & 4.743        & 4.099        \\
DFT                                      & 2.427        & 1.781        \\
\hline \hline
\end{tabular}
\caption{\label{tab:Gaps} 
Direct and indirect QP and optical gaps for the monolayer SrTiO$_3$, including the exciton binding energies $E_{xc}$ of the first excitonic state in parentheses. Both $G_0W_0^{full}$ and $G_0W_0^{diag}$ predicted QP gaps are shown. The optical gaps and $E_{xc}$ are determined by the solution of BSE starting from $G_0W_0^{full}$ predicted QP eigenstates.}
\end{table}
\newline The experimental synthesis of freestanding TMO perovskites monolayers opens therefore the possibility of extending these analysis to SrTiO$_3$. Nevertheless, to the best of our knowledge, the studies of excitonic effect on this compound have been restricted to the bulk phase and thin films that do not reach the monolayer limit.
\newline\newline In this work we present a first-principles study of the optical spectra and the underlying excitonic transitions of monolayer SrTiO$_3$, determined through the solution of the Bethe-Salpeter equation (BSE), where the quasiparticle (QP) eigenstates and screened Coulomb 
interactions are computed by the GW method.
Our calculations indicate that DFT predicts incorrect hybridizations  between Ti-$d$ and  O-2$p$ orbitals for the lower conduction bands and thus yields incorrect band characters.
This leads to a severe failure for the diagonal G$_0$W$_0$ method when taking DFT one-electron energies and orbitals as a starting point. 
The correct band dispersions are restored only by  taking into account the off-diagonal elements of the self-energy.
We find that excitonic effects significantly alter the optical spectra.
In addition to the spectral weight transfer visible also in bulk, new bound exciton peaks appear in monolayer SrTiO$_3$.
The origin of these excitonic peaks is clarified by analyzing the coupling components of the BSE eigenstates associated to these peaks.
Moreover, we show that reduced dimensionality effects cause a considerable enhancement of exciton binding energies,
yielding a binding energy as large as $0.93$ eV for the first exciton at the optical direct gap.
\newline Our calculations were performed using the VASP software,~\cite{ PhysRevB.54.11169,KRESSE199615} using a $24\times24\times1$ k-point mesh, a 600 eV cutoff energy for the wavefunctions and a 325 eV cutoff energy for the response functions.
Finite basis errors on the QP energies were corrected by introducing a scissor operator determined using the YAMBO software.~\cite{MARINI20091392,Sangalli2019}
\newline The geometry of monolayer SrTiO$_3$ (001) was determined through a structural optimization starting from the relaxed cubic bulk phase.
The optimization results in distortions along the z axis (see Fig.~\ref{fig:QP_KS_BS}b), in agreement with previous works~\cite{RelaxedPan2021,RelaxedXiao2021}: The Ti and Sr atoms are displaced toward the inner side, with a larger displacement associated to the Sr atom, corresponding to the C4v symmetry group. The distortions give rise to a polarization perpendicular to the monolayer plane~\cite{RelaxedXiao2021} (see Fig.~\ref{fig:QP_KS_BS}), which was calculated to be 0.069 $|e|$\r{A} using the Berry Phase method.~\cite{PhysRevB.47.1651} This value is consistent with the one estimated by the approximated expression $\textbf{P}_{tot} = \sum_i \Delta R_i Z_i^*$ (where $\Delta R_i$ are the atomic displacements and $Z_i$ the Born effective charges) which gives 0.062 $|e|$\r{A}.

\begin{figure*}[tb]
	\includegraphics[width=0.9\linewidth]{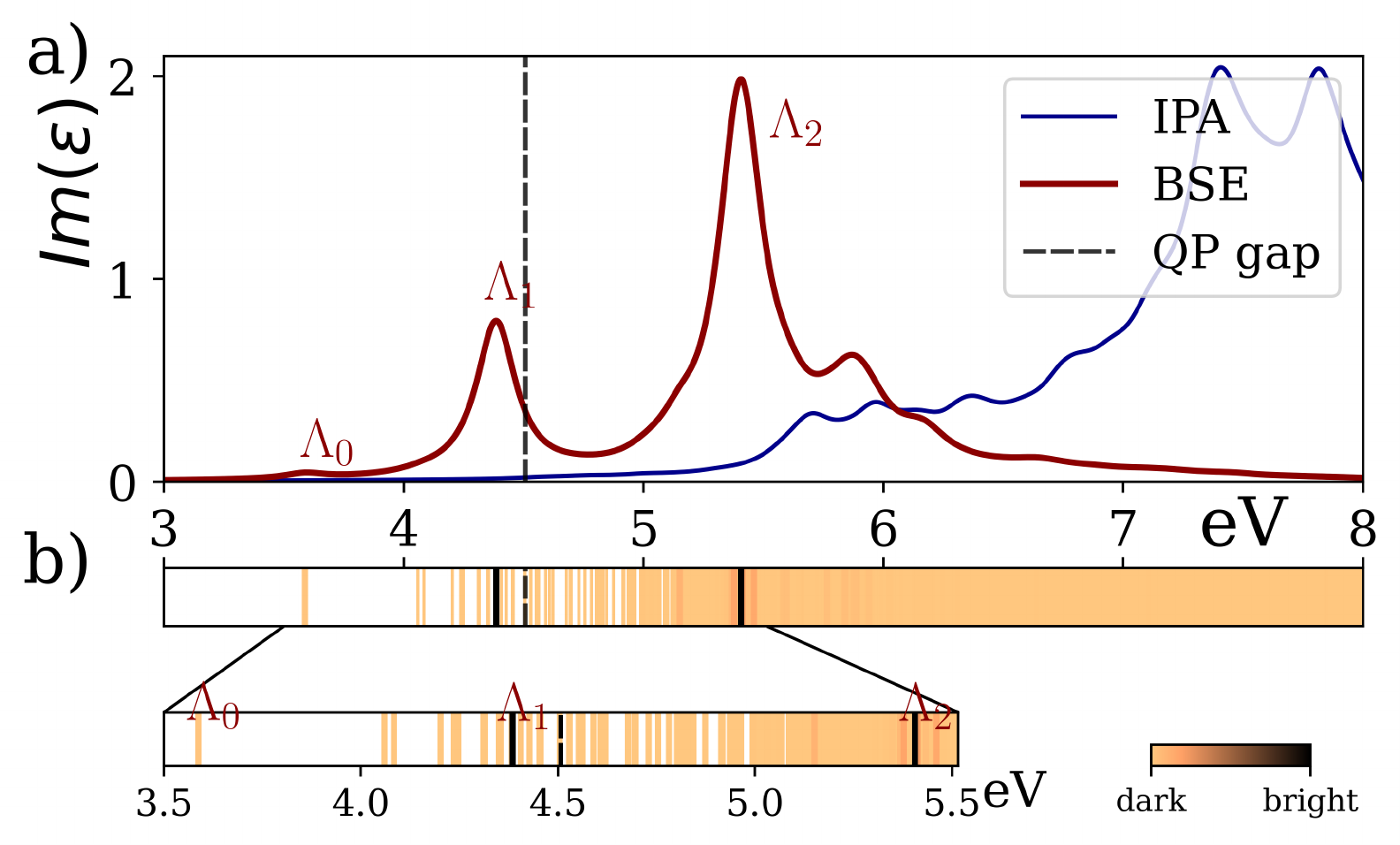}
	\caption{Imaginary part of dielectric functions with excitonic effects (BSE) and in the independent particle approximation (IPA).
The vertical dashed line represents the fundamental direct gap. The BSE eigenvalue spectrum is given in b), with an insert zooming over the bound exciton region.
The color coding indicates the associated oscillator strength with the maximum of the scale being set to that of $\Lambda_2$.}
\label{fig:spectra_barcode}
\end{figure*}
\paragraph{Electronic properties}
In order to obtain accurate optical properties, an accurate description of the QP band structures is indispensable~\cite{Review_RP}.
Here, the QP band structures were computed using a many-body $G_0W_0$ method~\cite{PhysRevB.34.5390,PhysRevB.74.035101,PhysRevB.75.235102}
with the screened interaction $W$ calculated by the random phase approximation (RPA) using DFT one-electron energies and orbitals.
Two different $G_0W_0$ schemes are discussed through this work.
The first is the widely used one-shot approach based on the diagonal approximation of the self-energy $\Sigma$ (referred to as $G_0W_0^{diag}$),
and the other one is the $G_0W_0^{full}$ approach where eigenvalues and orbitals are determined by diagonalizing an Hamiltonian constructed from the full dynamical self-energy matrix~\cite{PhysRevLett.99.246403,PhysRevLett.96.226402,PhysRevLett.93.126406,Reining-GW}. 
Remarkably, the QP corrections for monolayer SrTiO$_3$ are not just rigid energy shifts, but rather prove to be k-point- and band- dependent.
As displayed in Fig.~\ref{fig:QP_KS_BS}, the DFT bandstructure shows markedly different conduction bands dispersion with respect to $G_0W_0^{full}$ predicted ones.
\begin{figure*}[tb]
	\includegraphics[width=0.9\linewidth]{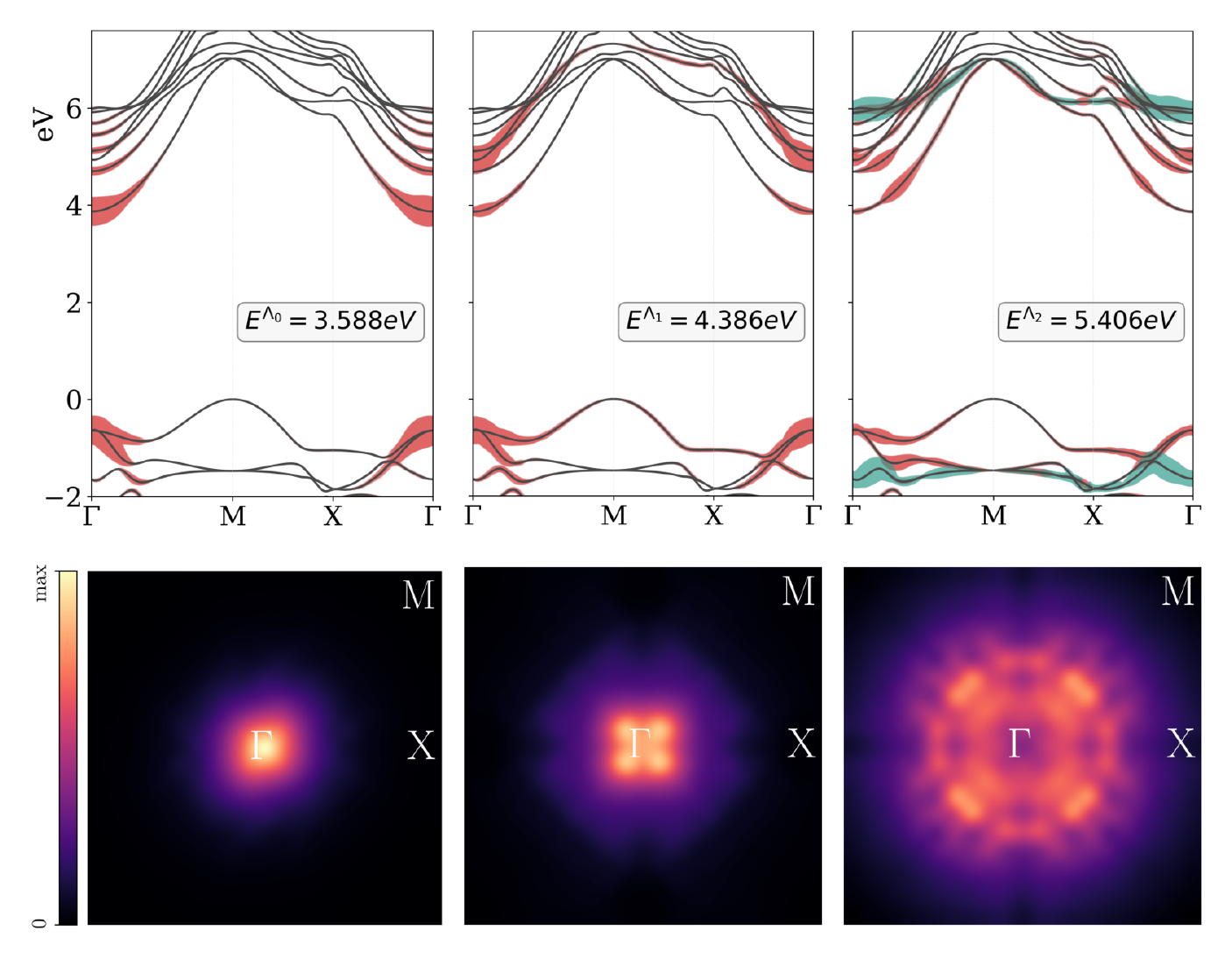}
	\caption{Upper panels: fatband analysis on the optical transitions associated with $\Lambda_0$, $\Lambda_1$ and $\Lambda_2$ states.
The fatness of the bands is proportional to the square of the amplitude of the electron-hole coupling coefficients
$|A^\Lambda_{\textbf{k} v c}|^2$ with $v$, $c$, $\textbf{k}$ and $\Lambda$ denoting the valence band index, conduction band index, $k$-point and BSE eigenvalue, respectively.
For $\Lambda_2$, the $\text{O-}p \rightarrow \text{Ti-}d_{yz}+\text{Ti-}d_{xz}$ optical transitions, discussed in the text, are highlighted by a light blue color.
	Lower panels: Distribution of the BSE eigenvectors in the Brillouin zone for the corresponding BSE eigenvalues.
The color coding denotes $\sum_{v,c} |A^\Lambda_{\textbf{k} v c}|^2$.}
\label{fig:fatband}
\end{figure*}
This difference is connected to an inadequate description of orbital characters by DFT. 
In the DFT bandstructure (grey lines in Fig.~\ref{fig:QP_KS_BS}a)  the conduction manifolds are dominated by Ti-$d$ states. 
By contrast, the conduction manifolds predicted by $G_0W_0^{full}$ exhibit a sizable hybridization between Ti-$d$ and Sr and O-$p$ states. In particular, the lowest conduction band possesses a considerable mixing of Sr states (up to $\sim40\%$) and O-$p$ (up to $\sim20\%$) along $\Gamma$-M and  X-$\Gamma$. 
The valence bands characters are unchanged between $G_0W_0^{full}$ and DFT and are clearly contributed by the O-$p$ states.
Moreover the QP valence bands can be successfully modeled using a typical scissor plus stretching correction.
It is instructive to compare these results with the widely used diagonal $G_0W_0$ approximation.
Since the $G_0W_0^{diag}$ method retains a pronounced starting point dependence~\cite{RevModPhys.74.601,10.3389/fchem.2019.00377,PhysRevLett.93.249701,PhysRevB.76.115109,doi:10.1063/1.4863502},
it is not so surprising that $G_0W_0^{diag}$ on top of the incorrect DFT band characters results in unphysical band dispersions and multiple crossings long X-$\Gamma$ among the lower conduction bands (see Fig.~\ref{fig:QP_KS_BS}a).
The failure of $G_0W_0^{diag}$ originates from the neglect of off-diagonal matrix elements of the self-energy, a coupling that can hybridize the single particle states.
Indeed, using the full self-energy operator correctly couples the single particle Kohn-Sham orbitals and restores the correct hybridizations, thus avoiding the multiple band crossings (see Fig.~\ref{fig:QP_KS_BS}a).
Therefore, going beyond the diagonal approximation for the self-energy is paramount for a correct description of the band dispersions of monolayer SrTiO$_3$.
We note in passing that similar behaviours have been observed in bulk topological insulators~\cite{PhysRevResearch.2.043105,PhysRevB.100.155147}
(in particular, Aguilera and coworkers~\cite{PhysRevB.88.045206} noted how in Sb$_2$Tb$_3$ and Bi$_2$Tb$_3$ unphysical band dispersions may arise for orbitals strongly affected by $GW$ corrections) and materials with strong $p$-$d$ hybridizations~\cite{Nabok2021,PhysRevB.74.245125,PhysRevB.84.115106,PhysRevLett.99.266402}.
\newline The fundamental quasiparticle bandgaps of monolayer SrTiO$_3$ are summarized in Table~\ref{tab:Gaps}. The DFT fundamental bandgap is indirect, with the valence band maximum (VBM) at M and the conduction band minimum (CBM) at $\Gamma$, while the DFT direct gap is defined at $\Gamma$.
Despite the strong effect  of QP corrections, the $G_0W_0^{full}$ indirect and direct gaps are opened at the same \textit{k}-points ($\Gamma-\text{M}$ and $\Gamma$), and are equal to $4.514$ eV and $3.870$ eV respectively.
\paragraph{Optical and excitonic properties}
The optical response of monolayer SrTiO$_3$ is dominated by excitonic effects.
The direct optical gap is  associated to a large excitonic binding energy of $0.93$ eV (see Table~\ref{tab:Gaps}). This value is much larger than the one estimated for the bulk phase ($\sim 0.205 \,–\, 0.240$ eV~\cite{PhysRevMaterials.5.074601,PhysRevB.87.235102,PhysRevMaterials.3.065004}). This is a typical consequence of the screening environment of 2D materials~\cite{PhysRevLett.113.076802,PhysRevB.84.085406,PhysRevB.88.245309}.
\newline The BSE predicted imaginary part of the dielectric function in Fig.~\ref{fig:spectra_barcode} 
is dominated by two very intense and narrow peaks.
This is in marked contrast to the one computed from the independent particle approximation, which exhibits a long absorption tail.
The first narrow peak is located below the QP direct gap and is determined by the excitonic state $\Lambda_1$. A low-intensity feature is visible at the optical direct gap and is associated to the lowest bound exciton $\Lambda_0$, with a considerable redshift at the onset energy of $\sim 1.0$ eV.
$\Lambda_0$ is related to a bright exciton, albeit with a very weak oscillator strength (less than $5\%$ of $\Lambda_1$).
The continuum region displays a single prominent structure, in the form of sharp peak (i.e. $\Lambda_2$ with strongest oscillator strength) plus a shoulder. 
\newline Now, we turn to the analysis of the fine structure of bound excitons in Fig.~\ref{fig:fatband}. 
The lowest state $\Lambda_0$ is doubly degenerate and weakly optically active, with a modest oscillator strength. 
The contributions to the excitonic wavefunction in reciprocal space $A_{\textbf{k} v c}^{\Lambda_0}$ are predominantly localized at $\Gamma$ and originate from optical transitions from the valence O-p states the CBM.
The low oscillator strength can be explained in terms of orbital characters of the CBM. Specifically, the conduction band minimum retains a substantial O-$p$ and Sr hybridization; in particular the Sr hybridization at $\Gamma$ is mainly composed by Sr-$s$ and Sr-$p$ characters. This is in turn associated to a partial suppression of the optical  matrix elements between the valence O-$p$ states and the CBM.
\newline The $\Lambda_1$ exciton is strongly bound, with an exciton binding energy of $1.31$ eV. We note that the binding energy of the $\Lambda_1$ exciton here was determined as the difference between the BSE eigenvalue and the interband transition with the strongest contribution in the excitonic eigenstate, following Refs.~\cite{PhysRevB.100.125413,Antonius2018}
The major $A^{\Lambda_1}_{\textbf{k} v c}$ terms correspond to interband transitions localized near $\Gamma$ from the two highest valence bands formed by O-$p_x/p_y$ states (doubly degenerate at $\Gamma$), to the second conduction band of Ti-$d_{xy}$ character (see Fig.~\ref{fig:fatband}). 
\newline The BSE eigenstate $A^{\Lambda_2}_{\textbf{k} v c}$ in the continuum mixes the (previously cited) O-$p_x/p_y \rightarrow \text{Ti-}d_{xy}$ optical transitions with a second excitation channel (highlighted with a different color in Fig.~\ref{fig:fatband}),  from the O-$p_z$ valence states to the non-dispersive conduction bands in the regions around $\Gamma-\text{X}$ and $\Gamma-\text{M}$ at $\sim~6.0$ eV, dominated by Ti-$d_{yz} / \text{Ti-}d_{xz}$ states (with a negligible hybridization with other states, less than $12\%$).
\newline The high intensity of the $\Lambda_2$ feature can be associated to the localization of the Ti-$d$ states in the non-dispersive regions at $\sim~6.0$ eV. We note moreover that all O-$p$ valence orbitals involved in $\Lambda_1$ and $\Lambda_2$ originate from the oxygen atoms situated in the Ti plane.  
Due to particular screening environment of 2D compounds, the orbitals perpendicular to the monolayer plane (e.g., the ones involved in the  $\text{O-}p_z \rightarrow \text{Ti-}d_{yz} / \text{Ti-}d_{xz}$ channel) experience a reduced screening with respect to the plane confined ones~\cite{PhysRevB.100.125413}; this effect concurs to explain the large $\sim 2.0-2.1$ eV redshift of the $\Lambda_2$ peak. 
\begin{figure}[bt]
	\includegraphics[width=0.9\linewidth]{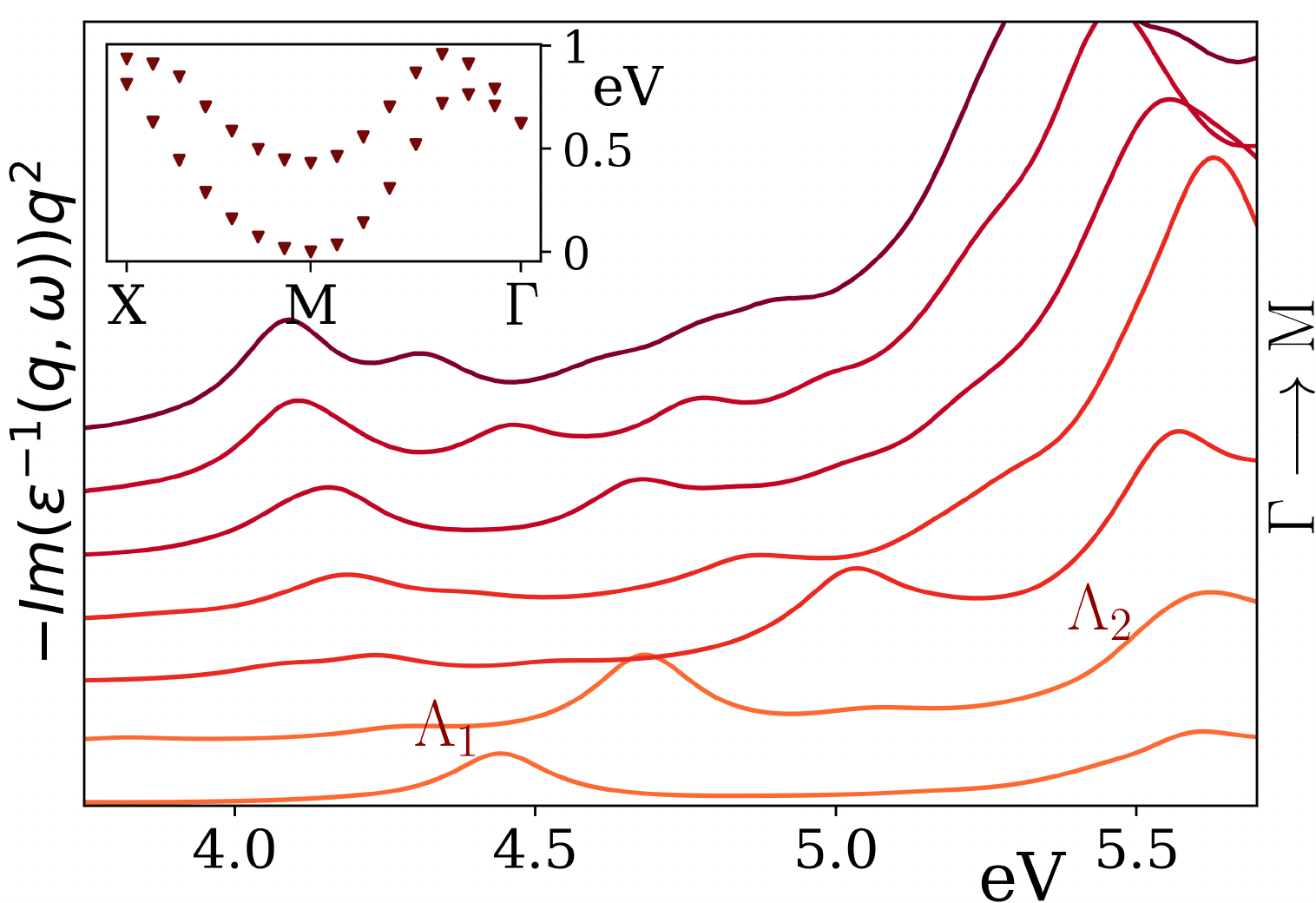}
	\caption{\label{fig:LOSS}  Loss  function for transferred momenta along the high-symmetry direction $\Gamma$-M, from $\textbf{q}=\text{M}/8$ to $\textbf{q}=\text{M}$. 
Each curve is multiplied by $q^2$, following Cudazzo and coworkers~\cite{ExcBS_Cudazzo,ExcBS_Bonacci}. The insert shows the excitonic band structure for the lowest two excitons along $\text{X}-\text{M}-\Gamma$; the zero energy is set at the eigenvalues minimum at $\text{M}$.}
\end{figure}
\newline\newline Next, we turn to the investigation of the excitonic dispersion at finite $\textbf{q}$, that is beyond the optical limit. This allows us to further characterize and discriminate the the excitonic properties in 2D systems~\cite{ExcBS_Cudazzo,ExcBS_Bonacci,ExcBS_Fugallo,ExcBs_Qiu,ExcBS_Sponza}.
The excitonic dispersion can be accessed experimentally by means of electron energy loss spectroscopy (EELS) or resonant inelastic x-ray spectroscopy (RIXS)~\cite{ExcBS_Cudazzo}. In particular in the EELS technique the cross section depends on the Loss function $L(\textbf{q},\omega) = - Im \left( \epsilon^{-1}(\textbf{q},\omega)  \right)$.
\newline Our computed Loss functions for various $\textbf{q}$ are plotted in Fig.~\ref{fig:LOSS} along $\Gamma-M$ (corresponding to the indirect gap direction).
\newline It's interesting to find that the doubly degenerate lowest excitonic state $\Lambda_0$, that yields a low-intensity feature for $q \rightarrow 0$, becomes completely optically inactive along $\Gamma$-M.
Furthermore, the analysis of exciton dispersions (insert in Fig.~\ref{fig:LOSS}) shows that the double-degeneracy is splitted away from $\Gamma$, and the two resulting excitonic bands reach their minimum at $\textbf{q}=\text{M}$, in correspondence of the indirect QP bandgap. 
The lowest (dark) excitonic band shows a parabolic dispersion around $M$, with an associated binding energy of 0.98 eV at $M$.
\newline Upon increasing momentum transfer, the peak associated with the $\Lambda_1$ state disperses to higher energies and progressively merges with the high-intensity structure at $\sim5.5$ eV (identifiable with the $\Lambda_2$ transition).
At large $\textbf{q}$ a new feature appears at transition energies around $4.1$ eV, which originates from the interband transitions from the three highest valence bands to the lowest conduction band. 
In particular, for $\textbf{q}=M$ a non-negligible contribution to the BSE eigenstate  $A_{v c \textbf{k}}(\textbf{q}=\text{M})$ (up to 30\% of the total spectral weight) is obtained by the transitions from valence $\text{O-}p_z$ states, which are also involved also in the $\Lambda_2$ excitonic transition.
\newline\newline In summary, we have investigated the quasiparticle and excitonic properties of freestanding monolayer SrTiO$_3$, using an \textit{ab initio} approach based on many body perturbation GW theory and Bethe-Salpeter equation.
We demonstrate that the inclusion of off-diagonal self-energy matrix elements in the G$_0$W$_0$ scheme is crucial to correctly describe the strong hybridization of the lower conduction bands (which is wrongly accounted for by DFT) and hence prevents the appearance of unphysical dispersions.
\newline The excitonic properties both in the optical limit $\textbf{q}\rightarrow0$ and for finite momenta have been studied.
We find that the spectra at $\textbf{q}\rightarrow0$ is dominated by excitonic effects, with a large binding energy of $\sim0.93$ eV at the direct optical gap.
The analysis of the BSE coupling components shows that the most intense peaks are originated from the O-$p\rightarrow $Ti-$d$ interband transitions, in conformity with the bulk description.
In particular, the transitions from the in-plane O-$p_x$/$p_y$ orbitals to Ti-$d_{xy}$ orbitals
and from out-of-plane O-$p_z$ orbitals to Ti-$d_{xz}/d_{yz}$ orbitals form separate excitation channels, which allows us to differentiate the two peaks.
At finite $\textbf{q}$, the lowest exciton becomes inactive with a parabolic excitonic dispersion around the transition minimum at $\textbf{q}=\text{M}$, an energy lower than the direct optical gap. 
\newline Our work opens a path in exploring the excitonic properties of 2D TMO perovskites and sets the basis fo on-going and future experimental and computational studies on ultrathin materials.
\newline\newline
\begin{acknowledgement}
\newline The authors thank G. Kresse, M. Marsili and J. He for fruitful discussions.
The computational results have been achieved using the Vienna Scientific Cluster (VSC) and the Galileo100 cluster (CINECA, LIMIT project).
\end{acknowledgement}
\bibliography{main}

\end{document}